\documentclass[preprint,aps]{revtex4}
\usepackage{graphicx}% Include figure files
\usepackage{dcolumn}% Align table columns on decimal point
\usepackage{bm}% bold math
\usepackage{color}
\begin{document}

%\begin{flushleft}
%prepared for \textit{Phys. Rev. Lett.; Version \#7}
%\end{flushleft}

\title{Induced magnetism of carbon atoms at the graphene/Ni(111) interface}

\author{M. Weser,$^1$ Y. Rehder,$^1$ K. Horn,$^1$ M. Sicot,$^2$ M. Fonin,$^2$ A. B. Preobrajenski,$^3$ E. N. Voloshina,$^4$ E. Goering,$^5$ and Yu. S. Dedkov$^{1,}$\footnote{Corresponding author. E-mail: dedkov@fhi-berlin.mpg.de}}
\affiliation{$^1$Fritz-Haber Institut der Max-Planck Gesellschaft, 14195 Berlin, Germany}
\affiliation{\mbox{$^2$Fachbereich Physik, Universit\"at Konstanz, 78457 Konstanz, Germany}}
\affiliation{$^3$MAX-lab, Lund University, 22100 Lund, Sweden}
\affiliation{$^4$Institut f\"ur Chemie und Biochemie - Physikalische und Theoretische Chemie, Freie Universit\"at Berlin, Takustra\ss e 3, 14195 Berlin, Germany}
\affiliation{\mbox{$^5$Max-Planck-Institut f\"ur Metallforschung, Heisenbergstrasse 3, 70569 Stuttgart, Germany}}

\date{\today}

\begin{abstract}
We report an element-specific investigation of electronic and magnetic properties of the graphene/Ni(111) system. Using x-ray magnetic circular dichroism, the occurrence of an induced magnetism of the carbon atoms in the graphene layer is observed. We attribute this magnetic moment to the strong hybridization between C $\pi$ and Ni $3d$ valence band states. The net magnetic moment of carbon in the graphene layer is estimated to be in the range of $0.05-0.1\,\mu_B$ per atom.
\end{abstract}


\maketitle

Among the allotropes of carbon, graphene, the two-dimensional hexagonally coordinated $sp^2$-bonded form has received enormous attention in recent years, because of its unique physical properties~\cite{Geim:2007a,Neto:2009}. A long electronic mean free path~\cite{Geim:2007a} and negligible spin-orbit coupling in graphene~\cite{Min:2006} leading to large spin relaxation times render this material ideal for applications based on ballistic transport like the spin filed effect transistor~\cite{Datta:1990}. In fact, graphene-based spin electronic devices possess a tremendous potential for high-density non-volatile memories, reconfigurable electronic devices and, possibly, solid-state quantum computing elements~\cite{Falko:2007,Rycerz:2007,Trauzettel:2007}. In particular, recent theoretical study~\cite{Karpan:2007} suggested to use graphene as a building block in spin-filtering devices, where the perfect spin-filtering was calculated for sandwiches composed of Ni(111) electrodes separated by graphene layers. As the spin dependent transport depends on the interface quality, the graphene/Ni(111) system is of special interest providing an ideal interface from a structural point of view due to the small mismatch of only 1.3\% between the two lattices~\cite{Gamo:1997}. Besides the fact that graphene/Ni-based structures possibly give an alternative route to devices based on tunneling magnetoresistance, graphene/Ni bilayer might also help to overcome some of the major obstacles in the field of spintronics, such as low spin injection efficiency. However, prior to being able to implement graphene/ferromagnet systems in any kind of spintronic device, a study of the electronic, magnetic and interfacial properties has to be performed.

Here we report on the investigation of the electronic and magnetic properties of the graphene/Ni(111) interface. By using x-ray magnetic circular dichroism (XMCD) we were able to identify the magnetic carbon moment in the graphene single layer on Ni(111). Our experimental results are discussed in the light of recent density-functional theory calculations and previous results on the observation of induced magnetism in non-magnetic materials.

Experiments were performed at the D1011 beamline of the MAX-Laboratory Synchrotron Facility (Lund, Sweden). The base pressure in the experimental station is $5\times10^{-11}$\,mbar, rising to $3\times10^{-10}$\,mbar during metal evaporation. An ordered graphene overlayer on Ni(111) was prepared as described elsewhere~\cite{Dedkov:2008,Dedkov:2008a,Dedkov:2008b}. A low-energy electron diffraction (LEED) pattern of the graphene/Ni(111) system revealed a well-ordered $p(1\times1)$ pattern, without any additional reflexes, as expected from the small lattice mismatch of only 1.3\% [Fig.\,1(a)]. STM measurements (performed in a separate experimental station under the same experimental conditions) indicate that a high quality epitaxial graphene layer was formed on the Ni(111) substrate [Fig.\,1(a)]~\cite{Dedkov:2008,Dedkov:2008a}. X-ray photoelectron spectroscopy (XPS) and angle-resolved photoemission (ARPES) (angular resolution of $0.5^\circ$) measurements were performed with a SCIENTA200 energy analyzer with an energy resolution of 100\,meV. XAS and XMCD spectra were collected at the Ni $L_{2,3}$ and C $K$ absorption edges in partial (repulsive potential $U=-100$\,V) and total electron yield modes (PEY and TEY) with an energy resolution of 100\,meV. Magnetic dichroism spectra were obtained with circularly polarized light (the degree of polarization was $P=0.75$) at different angles $\alpha$ in the remanence magnetic state of the graphene/Ni(111) system after applying of magnetic field of 500\,Oe along the $<1\bar{1}0>$ easy magnetization axis of the Ni(111) thin film. All spectra were recorded at room temperature.

Fig.\,1 (b) and (c) show the XPS C $1s$ ($h\nu=450$\,eV) and normal emission valence band spectra ($h\nu=65$\,eV), respectively, of the graphene/Ni(111) system. The C $1s$ photoelectron spectrum contains only one component and is similar to that reported for single crystalline graphite~\cite{Prince:2000,Lizzit:2007}, graphene/Pt(111), and graphene/Ir(111)~\cite{Preobrajenski:2008}.

The normal emission valence band spectra of single-crystalline graphite and graphene/Ni(111) presented in Fig.\,1 (c) were found to be in perfect agreement with previously published data~\cite{Dedkov:2008,Dedkov:2008a}. The shift to larger binding energy is different for $\sigma$ and $\pi$ valence band graphene-derived states. This behavior can be explained by the different strength of hybridization between these states and Ni $3d$ valence band states which is larger for the out-of-plane oriented $\pi$ states compared with the one for the in-plane oriented $\sigma$ states of the graphene layer. The observed binding energy difference of $\approx2.4$\,eV of the $\pi$ states ($\approx1$\,eV for the $\sigma$ states) for graphite and graphene on Ni(111) is in good agreement with previously reported experimental and theoretical values~\cite{Dedkov:2008a,Bertoni:2005}, reflecting a strong hybridization between these states and spin-polarized Ni $3d$ states.

In order to address the orbital orientation at the graphene/Ni interface, linearly polarized light was used and the sample orientation relative to the x-ray wave vector was varied~\cite{Stohr:1999b}. Moreover, the detection mode was chosen as PEY since it is more sensitive to the interface compared with TEY. The C $1s$ XAS spectra of the graphene/Ni(111) system were measured as a function of the angle, $\alpha$, between the direction of the incoming linearly polarized light and the surface of the sample, i.e. between the electrical vector of the light and the sample normal as depicted in Fig.\,2 (a).  The XAS spectrum of a graphite crystal measured at $\alpha=30^\circ$ is shown in the upper part of the figure and serves as a reference. Spectral features observed in the regions $283-289$\,eV and $289-315$\,eV are ascribed to C\,$1s\rightarrow\pi^*$ and C\,$1s\rightarrow\sigma^*$ transitions, respectively. The shape of XAS lines in both absorption regions is influenced by considerable excitonic effects through poor core-hole screening~\cite{Bruhwiler:1995,Ahuja:1996,Wessely:2005}. In the case of the graphene/Ni(111) system, the XAS C $1s\rightarrow\pi^*,\sigma^*$ spectrum shows considerable changes compared with the graphite spectrum, indicating a strong chemisorption. Similar effects but not as pronounced as in the present case, were recently observed for the graphene/Rh(111) and graphene/Ru(0001) interfaces~\cite{Preobrajenski:2008}. We interpret the broadening of the $\pi^*$ and $\sigma^*$ resonances as evidence for strong orbital hybridization and electron sharing at the graphene/Ni interface, indicating a strong delocalization of the corresponding core-excited state. A comparison of the present results for the graphene/Ni(111) system with those obtained earlier for graphene/Rh and graphene/Ru indicates the existence of a strong covalent interfacial bonding between carbon and Ni atoms. The small shoulder is also visible in the XAS spectra at 283.7\,eV of photon energy and can be associated with the lowering of the Fermi level induced by the charge transfer~\cite{Mele:1979,Liu:2004}. Since this feature does not demonstrate any dichroic signal (see below) it will not be discussed any further. 

The present XAS results can be analyzed on the basis of recent calculations of spin-resolved electronic structure and C $K$-edge electron energy loss spectra (EELS) for the graphene/Ni(111) interface~\cite{Bertoni:2005}. Our XAS spectra obtained at $\alpha=10^\circ$ and $\alpha=90^\circ$ correspond to the calculated EELS spectra for the scattering vector $\mathbf{q}$ perpendicular and parallel to the graphene layer, respectively. There is very good agreement between experimental XAS and calculated EELS spectra: (i) they posses the same angle (scattering vector) dependence and (ii) the spectral features in the experimental spectra are well reproduced in the theoretical ones. For example, two peaks in the XAS spectra at 285.5\,eV and 287.1\,eV of photon energy in the $1s\rightarrow\pi^*$ spectral region can be assigned to the double peak structure in the calculated EELS spectrum at 0.8\,eV and 3.0\,eV  above the Fermi level~\cite{Bertoni:2005}. In this case the first feature in the XAS spectrum corresponds to the transition of the electron from the $1s$ core level into the interface state above the Fermi level (around the $K$-point in the hexagonal Brillouin zone) which originates from C\,$p_z$--Ni\,$3d$ hybridization and corresponds to carbon atom C-$top$ and interface Ni atom antibonding state [for the structure of the graphene/Ni(111) system, see Fig.\,2(b,c)]. The second peak in the spectrum corresponds to the  transition of the electron from $1s$ core level on the interface state above the Fermi level (around the $M$-point in the hexagonal Brillouin zone) which originates from C\,$p_z$--Ni\,$p_x,p_y,3d$ hybridization and corresponds to a bonding between the two carbon atoms, C-$top$ and C-$fcc$, which involves the nickel atom at the interface. The theory also correctly describes the shape of the absorption spectra for the XAS C $1s\rightarrow\sigma^*$ transition~\cite{Bertoni:2005}.

The existence of a considerable magnetic interaction between the Ni substrate and the graphene film is proven by the analysis of both Ni $L_{2,3}$ and C $K$ XMCD spectra shown in Figure\,3. Our Ni $L_{2,3}$ XMCD spectrum of the graphene/Ni(111) system [Fig.\,3(a)] is in perfect agreement with previously published data~\cite{Srivastava:1998,Dhesi:1999,Nesvizhskii:2000}. The bulk values of the spin and orbital magnetic moments $\mu_S=0.69\,\mu_B$ and $\mu_L=0.07\,\mu_B$ of Ni calculated from the Ni $L_{2,3}$ TEY XAS spectra on the basis of sum-rules are in very good agreement with previously published experimental values~\cite{Baberschke:1996,Srivastava:1998} as well as with the spin-magnetic moment ($\mu_S=0.67\,\mu_B$) calculated for the graphene/Ni(111) system~\cite{Bertoni:2005}.

Fig.\,3 (b) depicts the dichroic signal at the C $K$ edge. This signal shows unambiguously that Ni induces a magnetic moment in the graphene layer. In order to increase the measured magnetic contrast at the $1s\rightarrow\pi^*$ absorption edge, these XMCD spectra were collected in the PEY mode at an angle of $\alpha=20^\circ$. The observed difference in the XAS spectra collected at this angle in Fig.\,2 and 3 is due to the different polarization of the light: it is linearly polarized in Fig.\,2, revealing a strong angular dependence of absorption due to the different orbital orientation in graphene, and circularly polarized in Fig.\,3, where both $1s\rightarrow\pi^*$ and $1s\rightarrow\sigma^*$ transitions are nearly equivalent. The C $K$ XMCD spectrum shows that the major magnetic response stems from transitions of the $1s$ electron into the $\pi^*$-states, while transitions into the $\sigma^*$-states yield practically no magnetic signal, indicating that only the C $2p_z$ orbitals which hybridize with the Ni $3d$ band are spin-polarized. The sharp structure at the $1s\rightarrow\pi^*$ absorption edge originates from the hybridized C\,$p_z$--Ni\,$3d$ and C\,$p_z$--Ni\,$p_x,p_y,3d$ states, see earlier discussion and \cite{Bertoni:2005}.

At the C $K$ edge, transitions occur from non-spin-orbit split $1s$ initial states to $2p$ final states. Thus the corresponding dichroic signal can only provide information on the orbital moment~\cite{Thole:1992,Carra:1993}. We conclude from the negative sign of the XMCD signal that the averaged orbital moment of carbon atoms, \textit{i.e.} averaged over all carbon positions in the graphene layer is aligned parallel to both, the spin and orbital moments of the nickel layer. It is noteworthy that the orientation of individual spin and orbital moments of both Ni and C at different sites can not be determined from the experimental XMCD data.

At this point of the analysis, one can ask for a quantitative value of the induced magnetic moment per C atom. However, due to the impossibility to extract the spin magnetic moment form the \textit{K} edge XMCD spectra, we apply a comparison with similar system to estimate it. Ferromagnetism of carbon in Fe/C multilayers was demonstrated and a magnetic moment of $0.05\,\mu_B$ was measured~\cite{Mertins:2004}. In these Fe/C multilayers, magnetism was shown to be related  to the hybridization of the Fe $3d$ orbitals and C $p_{z}$ orbitals exaclty like in the case of graphene/Ni(111). In addition, induced magnetism in carbon nanotubes in contact with a flat ferromagnetic Co substrate was demonstrated and a spin transfer of $0.1\,\mu_B$ per carbon atom was deduced~\cite{Cespedes:2004}. For all these reasons and considering these analogous systems, we estimate the induced magnetic moment for graphene on Ni(111) to be in the range of $0.05-0.1\,\mu_B$ per carbon atom.

\textit{In conclusion},  the electronic structure and magnetic properties of the graphene/Ni(111) interface were studied by means of x-ray absorption spectroscopy and magnetic circular dichroism at the Ni $L_{2,3}$ and C $K$ absorption edges. The XAS C $1s\rightarrow\pi^*,\sigma^*$ spectra show drastic changes with a variation of the angle between the electrical vector of the light and the surface normal that reflects the symmetry of the final states ($\sigma$ or $\pi$). Magnetic dichroism experiments reveal an induced magnetic moment of the $\pi$-electrons of carbon atoms in the graphene layer. This magnetic moment is due to the strong hybridization between C $\pi$ and Ni $3d$ valence band states. The magnetic moment of carbon atoms in the graphene layer was estimated to be in the range of $0.05-0.1\,\mu_B$ per carbon atom. Observed magnetism in graphene layer induced by a ferromagnetic substrate is of crucial importance for the design of new carbon-based spintronic devices.

\clearpage

{\bf Figure captions:}
\newline
\newline
{\bf Fig. 1.} (Color online) (a) STM and LEED characterization of the graphene layer on Ni(111). (b) C $1s$ photoemission spectrum of the graphene/Ni(111) system obtained with photon energy of 450\,eV. (c) Comparison of the normal emission valence band photoelectron spectra of the pure graphite crystal and graphene/Ni(111) collected at 65\,eV of photon energy. The main valence band-derived features of both systems are marked in the figure.
\newline
\newline
{\bf Fig. 2.} (Color online) (a) A series of the angle-dependent XAS spectra of the graphene/Ni(111) system (angle, $\alpha$, between direction of the light and the surface of the sample is marked on the left-hand side of the series). For comparison the XAS spectrum of bulk graphite measured at $\alpha=30^\circ$ is shown in the upper part of the plot. The inset shows the geometry of the XAS experiment. (b,c) Top and side view of the $top-fcc$ arrangement of carbon atoms in the graphene layer on Ni(111).
\newline
\newline
{\bf Fig. 3.} (Color online) XMCD spectra of the graphene/Ni(111) system: absorption spectra measured for two opposite orientations of the magnetization are shown in the upper part for the Ni $L_{2,3}$- (a) and C $K$-edges (b). The corresponding differences are shown in the lower part of the respective figures.

\clearpage
\begin{figure}[t]\center
\includegraphics[scale=0.55]{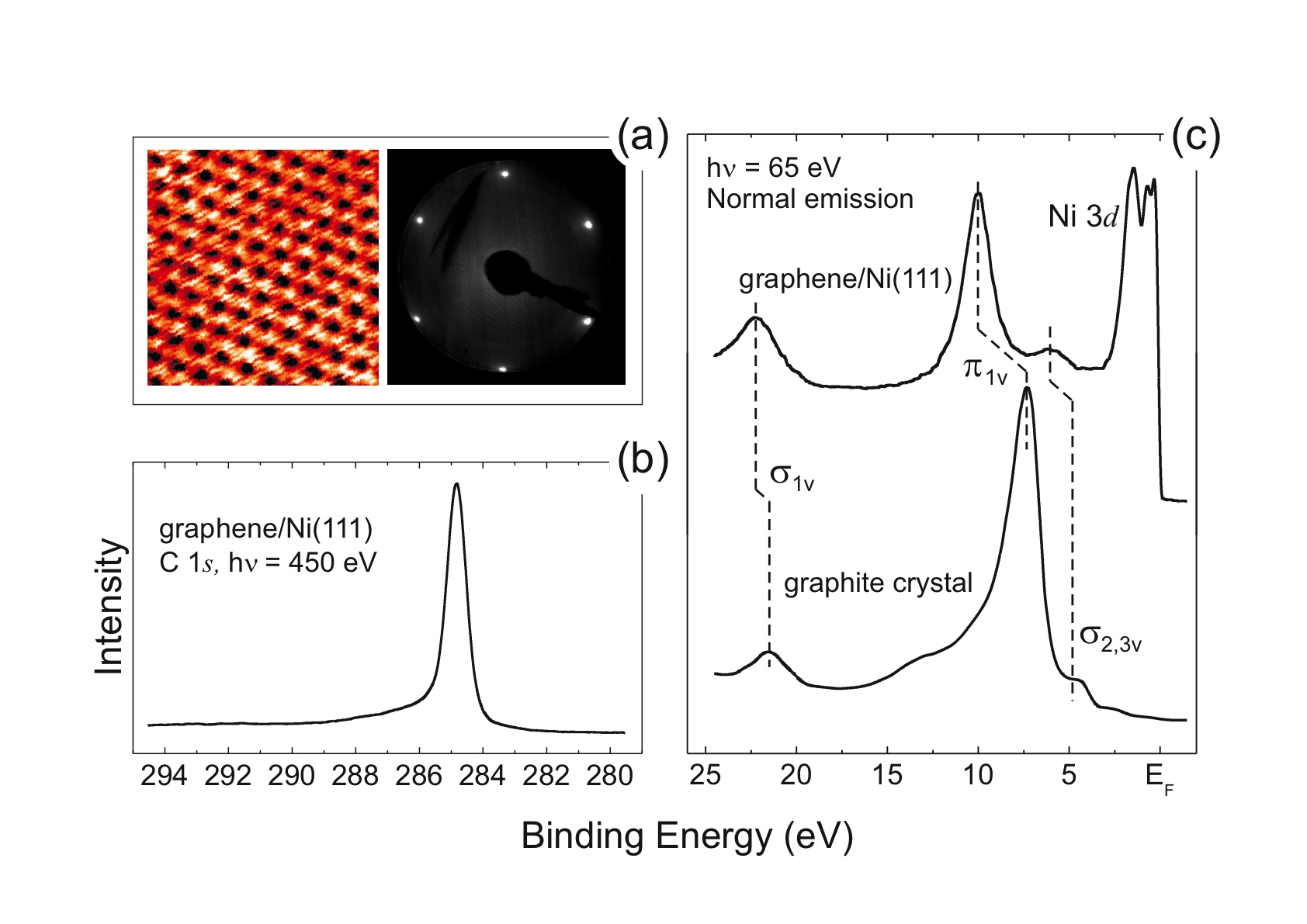}\\
\vspace{1cm}
\large \textbf{Fig.\,1.}
\end{figure}

\clearpage
\begin{figure}[t]\center
\includegraphics[scale=0.4]{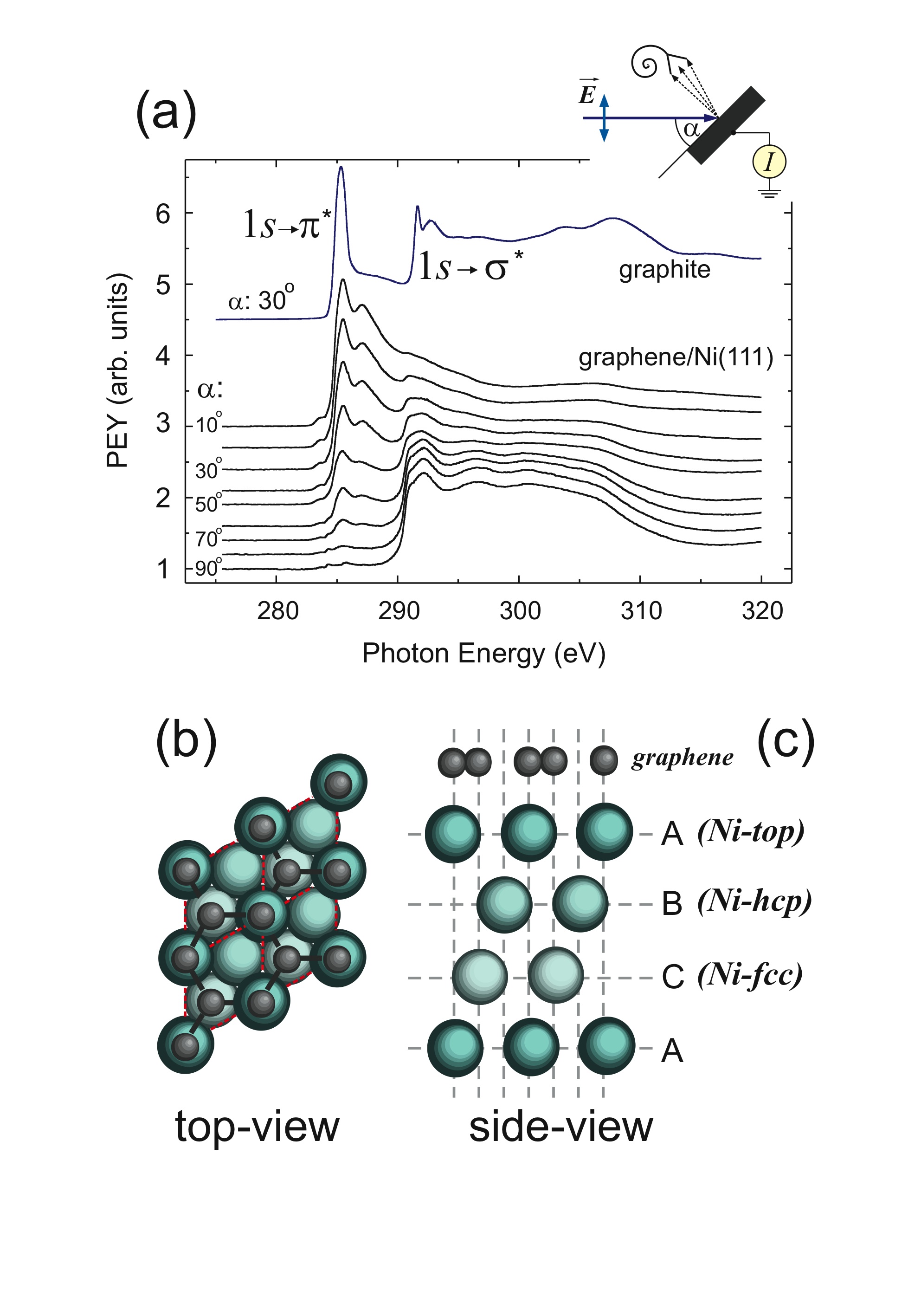}\\
\vspace{1cm}
\large \textbf{Fig.\,2.}
\end{figure}

\clearpage
\begin{figure}[t]\center
\includegraphics[scale=0.7]{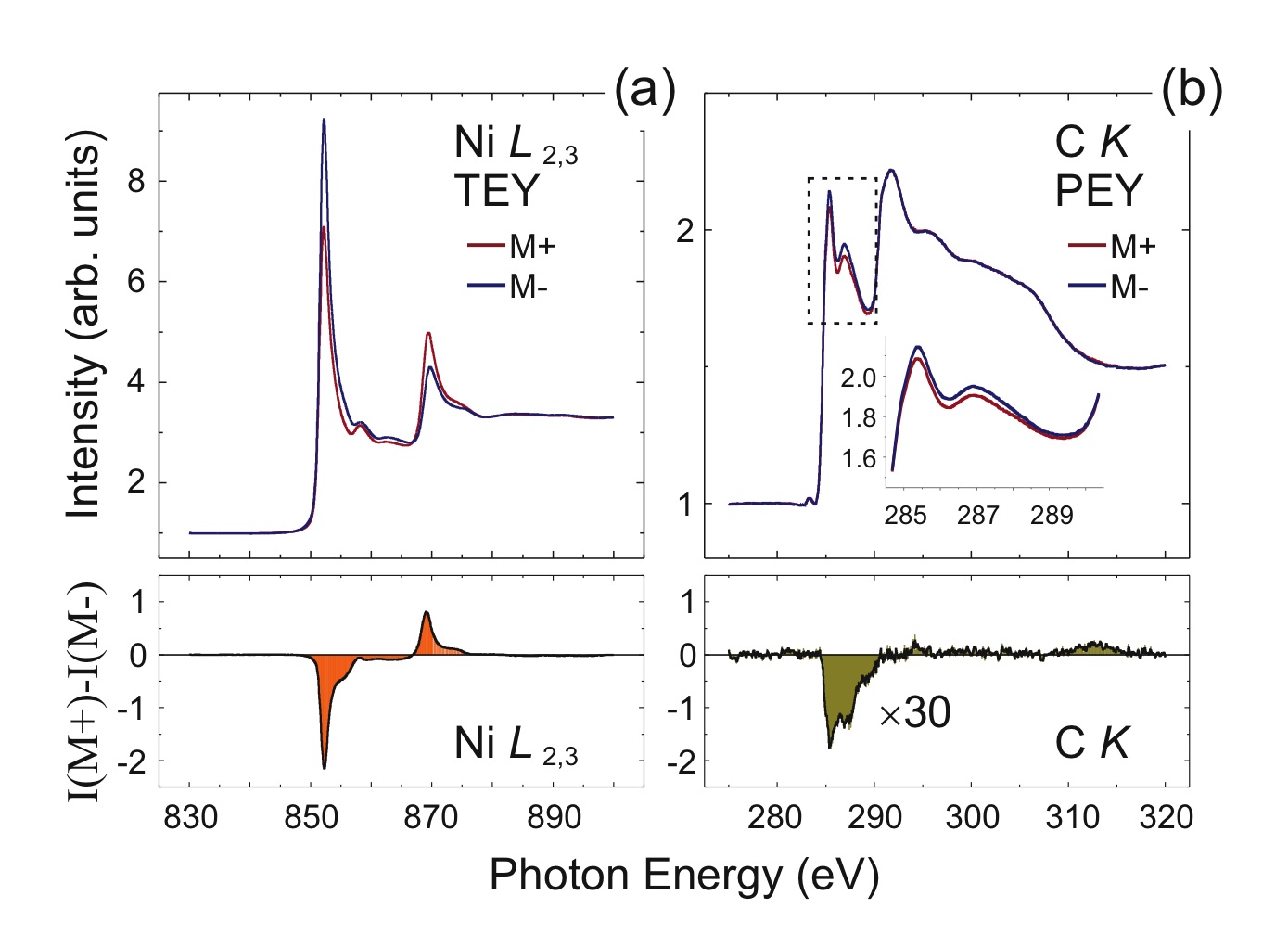}\\
\vspace{1cm}
\large \textbf{Fig.\,3.}
\end{figure}

\end{document}